\documentclass{iaus}
\usepackage{graphicx}
\usepackage{color}
\newcommand{\about}{$\sim\!\!$~}
\newcommand{\kms}{km s$^{-1}$}
\def\arcdeg{\hbox{$^\circ$}}

\usepackage[authoryear]{natbib}
\def\reff@jnl#1{{\rm#1\/}}
\def\apj{\reff@jnl{ApJ}}
\def\aj{\reff@jnl{AJ}}                  
\def\araa{\reff@jnl{ARA\&A}}            
\def\apj{\reff@jnl{ApJ}}                
\def\apjl{\reff@jnl{ApJ}}               
\def\apjs{\reff@jnl{ApJS}}              
\def\aap{\reff@jnl{A\&A}}               
\def\mnras{\reff@jnl{MNRAS}}            
\def\prd{\reff@jnl{Phys.Rev.D}}         
\def\prl{\reff@jnl{Phys.Rev.Lett}}      
\def\pasp{\reff@jnl{PASP}}              
\def\nat{\reff@jnl{Nature}}             
\def\iaucirc{\reff@jnl{IAU~Circ.}}%
\def\aaps{\reff@jnl{A\&AS}}%
\def\pasa{\reff@jnl{PASA}}%
\newcommand\actaa{\ref@jnl{Acta Astron.}}%
\newcommand\sovast{\ref@jnl{Soviet~Ast.}}%

\title[JD 11.~~Light Echoes] 
{Light Echoes of Historic Transients}

\author[A. Rest, B. Sinnott, D. L. Welch,  J. L. Prieto, and F. Bianco]   
{Armin Rest$^1$, B. Sinnott$^2$, D. L. Welch$^2$, J. L. Prieto$^3$ \and F. Bianco$^4$}

\affiliation{$^1$STScI, 3700 San Martin Dr., Baltimore, MD 21218, USA \\ 
email: {\tt arest@stsci.edu} \\[\affilskip]
$^2$Department of Physics and Astronomy, McMaster University, \\
Hamilton, Ontario L8S 4M1, Canada \\
$^3$Department of Astrophysical Sciences, Princeton University, \\
4 Ivy Lane, Princeton, NJ 08544, USA \\
$^4$Center for Cosmology and Particle Physics, New York University, \\
4 Washington Place, New York, NY 10003, USA}

\pubyear{2013}
\volume{296}  
\pagerange{119--126}
\jname{Supernova environmental impacts}
\editors{R. McCray, A. Ray, eds.}
\begin{document}

\maketitle

\begin{abstract}
Light echoes, light from a variable source scattered off dust, have
been observed for over a century. The recent discovery of light echoes
around centuries-old supernovae in the Milky Way and the Large
Magellanic Cloud have allowed the spectroscopic characterization of
these events, even without contemporaneous photometry and spectroscopy
using modern instrumentation. Here we review the recent scientific
advances using light echoes of ancient and historic transients, and
focus on our latest work on SN~1987A's and Eta~Carinae's light echoes.
\end{abstract}

\firstsection 
\section{Introduction}

Light echoes (LEs) arise when light from a transient or variable
source is scattered off circumstellar or interstellar dust, reaching
the observer after a time delay resulting from the longer path length
\citep[e.g.,][]{Couderc39,Chevalier86,Schaefer87a,Xu94,Sugerman03,Patat05}.
Over a century ago, in 1901, the first scattered LEs were discovered
around Nova Persei \citep{Ritchey01b,Ritchey01a,Ritchey02}.  They were
recognized as such shortly thereafter by \citet{Kapteyn02} and
\citet{Perrine03}. Since then, LEs have been observed around a wide
variety of objects: the Galactic Nova Sagittarii 1936 \citep{Swope40},
the eruptive variable V838 Monocerotis \citep{Bond03}, the Cepheid RS
Puppis \citep{Westerlund61,Havlen72}, the T Tauri star S~CrA
\citep{Ortiz10}, and the Herbig Ae/Be star R~CrA
\citep{Ortiz10}. Echoes have also been observed from extragalactic
SNe, with SN~1987A being the most famous case
\citep{Crotts88,Suntzeff88}, but also including SNe~1980K
\citep{Sugerman12}, 1991T \citep{Schmidt94,Sparks99}, 1993J
\citep{Sugerman02,Liu03}, 1995E \citep{Quinn06}, 1998bu
\citep{Garnavich01,Cappellaro01}, 2002hh \citep{Welch07,Otsuka12},
2003gd \citep{Sugerman05,VanDyk06,Otsuka12}, 2004et \citep{Otsuka12},
2006X \citep{Wang08,Crotts08}, 2006bc \citep{Gallagher11,Otsuka12},
2006gy \citep{Miller10}, 2007it \citep{Andrews11}, and 2008bk
\citep{VanDyk13}. All of the aforementioned LEs had the common
selection criterion that they were found serendipitously while the 
transient source was still bright.

Early on in the last century, \cite{Zwicky40} had the idea that it
might be possible to learn more about historical SNe by studying their
scattered LEs. However, the few dedicated surveys trying to implement
this idea for historic SNe
\citep{vandenBergh65b,vandenBergh65a,vandenBergh66,Boffi99} and novae
\citep{vandenBergh77,Schaefer88} were not successful.  With the
emergence of CCDs as astronomical detectors in combination with the
advancement in telescope technology that allowed to image larger
field-of-views, the wide-field time-domain surveys at the beginning of
this century significantly improved in depth and area.  These
improvements led to the first discoveries of LEs at angular distances
from the ancient transients too large to suggest an immediate association.
The i400-900 year-old LEs from three LMC supernovae were found by \cite{Rest05b} as part 
of the SuperMACHO survey \citep{Rest05a}.  Subsequent targeted searches in our 
Galaxy found LEs of Tycho's SN \citep{Rest07,Rest08b}, Cas~A \citep{Rest07,Rest08b,Krause08a}, and
$\eta$~Carinae \citep{Rest12_etac}.

\section{Light Echo Spectroscopy\label{sec:lespec}}

Spectroscopy of LEs allows the transient to be studied long
after it has already faded. The first LE spectrum was a 35~hour
exposure of one of Nova Persei 1901 LEs \citep{Perrine03}, confirming
that the nebulous moving features seen around Nova Persei were indeed
its echoes. LE spectra of SN~1987A were most similar to those of the
SN near maximum light \citep{Gouiffes88,Suntzeff88}.
Serendipitously, \cite{Schmidt94} found the spectrum of SN~1991T taken
750 days after maximum again similar to the one at peak, indicating that
at that time echoes from the SN at peak dominated the spectrum.

Spectroscopy of ancient SN LEs discovered in the LMC \citep{Rest05b}
resulted in the first opportunity to classify ancient transients long 
after their direct light had encountered Earth \citep{Rest08a}.  
The subsequent discovery and spectroscopy of LEs of Cas~A and
Tycho \citep{Rest07,Rest08b,Krause08a,Krause08b} allowed their
spectroscopic classification as a SN~IIb and normal SN~Ia,
respectively.

At first, it was widely believed that an observed LE spectrum would be
the lightcurve-weighted integration of the transients' individual
epochs \citep[e.g.,][]{Patat06,Rest08a,Krause08a}.  Such an
integration is equivalent to assuming that the dust filament is thick,
which is not always the
case\citep{Rest11_leprofile,Rest12b_lerev}. Spectral features which
persist for long periods of time during the evolution of an outburst
will be weighted much more strongly when scattered by a thick filament
than a thin filament.

In addition to spectroscopic classification, LEs also offer two more
exciting scientific opportunities: "3D spectroscopy and "spectroscopic
time series" of transients. Examples of realization of these techniques
are provide in the following two sections using SN~1987A and $\eta$~Carinae.

\subsection{3D Spectroscopy\label{sec:3Dspec}}

\begin{figure}[b]
\begin{center}
 \includegraphics[width=2.70in]{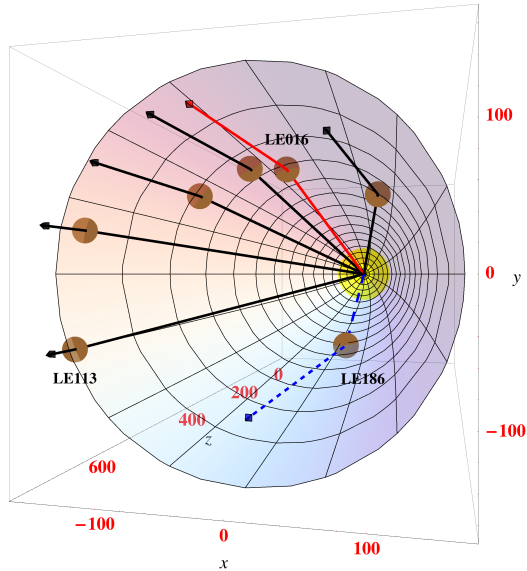} 
 \caption{Light paths and 3D scattering dust locations for seven LEs of 
 SN~1987A spectroscopically observed by \cite{Sinnott13}. Solid red 
 and dashed blue lines highlight the extreme north and south viewing 
 angles corresponding to LE016 and LE186 (see Figure~\ref{fig:87aLEspec}). 
 North is towards the positive y axis, east is towards
the negative x axis, and z is the distance in front of the SN. All
units are in light years. This figure is courtesy of \cite{Sinnott13}.}
   \label{fig:87a3D}
\end{center}
\end{figure}

LEs scattered by different dust structures offer an opportunity that
is unique in astronomy: probe the same object directly from different
directions. Fig.~\ref{fig:87a3D} shows the light paths of seven
SN~1987A LEs. 
By observing these LEs, SN~1987A can be analyzed as if the observers
are at different line-of-sights, allowing to directly compare
different hemispheres of one and the same object. This technique was
first applied to observe $\eta$~Carinae central star from different
directions using spectra of the reflection nebula
\citep{Boumis98,Smith03_eta}. 

\citet{Rest11_casaspec}
obtained spectra of three different Cas~A LEs, viewing the Cas~A
SN from very different lines-of-sight.  After accounting for the effects
of the scattering dust, they found that the He~I $\lambda 5876$ and
H$\alpha$ features of one LE are blue-shifted by an additional \about
4000~\kms\ relative to the other two LE spectra. X-ray and optical data of
the Cas~A remnant also show a Fe-rich outflow in the same direction
\citep{Burrows05, Wheeler08, Delaney10}. This indicates that
Cas~A was an intrinsically asymmetric SN.  The blue-shifted SN ejecta
is in the direction approximately opposite the motion of the resulting neutron star,
suggesting that the explosion mechanism that gave the neutron star its
kick affected the outer layers of the SN. This appears to be the first 
instance where the structure of the SN remnant can be directly associated
with asymmetry observed in the explosion itself.

\begin{figure}[b]
\begin{center}
 \includegraphics[width=3.4in]{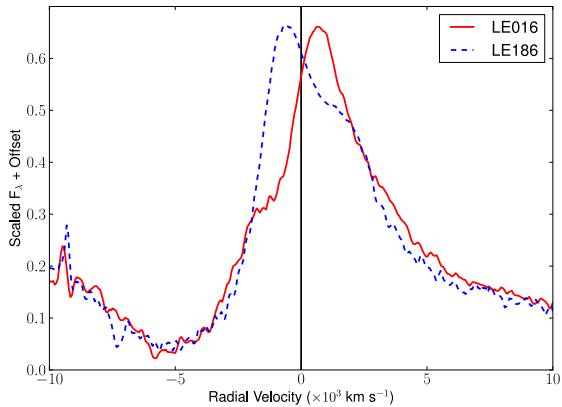} 
 \caption{Observed H$\alpha$ lines from LE016 and LE186. Emission peaks
have been interpolated with high-order polynomials. Spectra are scaled and
offset for comparison purposes, as well as smoothed with a boxcar of 3 pixels.
Although this plot does not take into account the important differences
in LE time-integrations between the spectra, it highlights the overall difference
in fine-structure in the two LE spectra from opposite PAs. Observing
H$\alpha$ profiles with opposite asymmetry structure at opposite PAs is surprising
considering the opening angle between the two LEs is $<40\arcdeg$. 
This figure and caption is courtesy of \cite{Sinnott13}.}
   \label{fig:87aLEspec}
\end{center}
\end{figure}

SN~1987A has a very rich set of LEs scattering off
circumstellar \citep{Crotts91,Crotts95,Sugerman05a,Sugerman05b} and
interstellar dust
\citep{Crotts88,Suntzeff88,Gouiffes88,Couch90,Xu94,Xu95}. The LEs 
shown in Figure~\ref{fig:87a3D} were spectroscopically observed
by \citet{Sinnott13}. After correcting for the effects of the
scattering dust, these spectra can be directly compared to a LE
spectrum constructed from the spectro-photometric library of SN~1987A 
\citep{87aspectra1,87aspectra2,87aspectra3,87aphoto_hamuy88,87aphoto_suntzeff88,87aspectra4,87aspectra5,Phillips88,Phillips90}.
 \citet{Sinnott13} find an excess in redshifted H$\alpha$ emission and
a blueshifted knee for the LE LE016 at position angle PA=16$\arcdeg$ (see red
line in Figure~\ref{fig:87aLEspec}).
Both asymmetry signatures disappear as the PA increases, and then reappear in 
the form of an excess in blueshifted H$\alpha$ emission and a redshifted knee at the 
opposite PA=186$\arcdeg$ LE, LE186 (see blue line in Figure~\ref{fig:87aLEspec}).

\begin{figure}[b]
\begin{center}
 \includegraphics[width=2.0in]{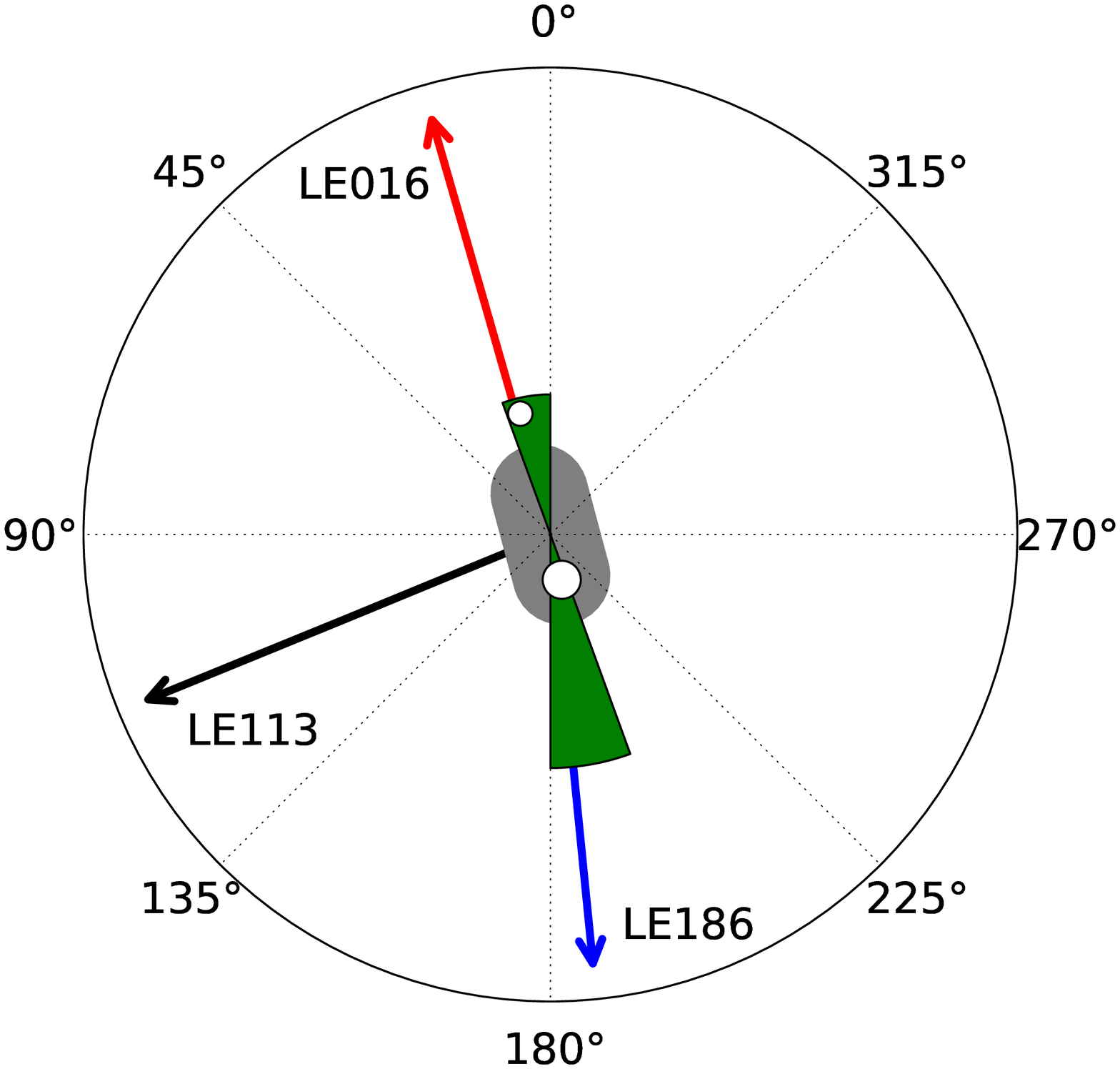} 
 \includegraphics[width=2.26in]{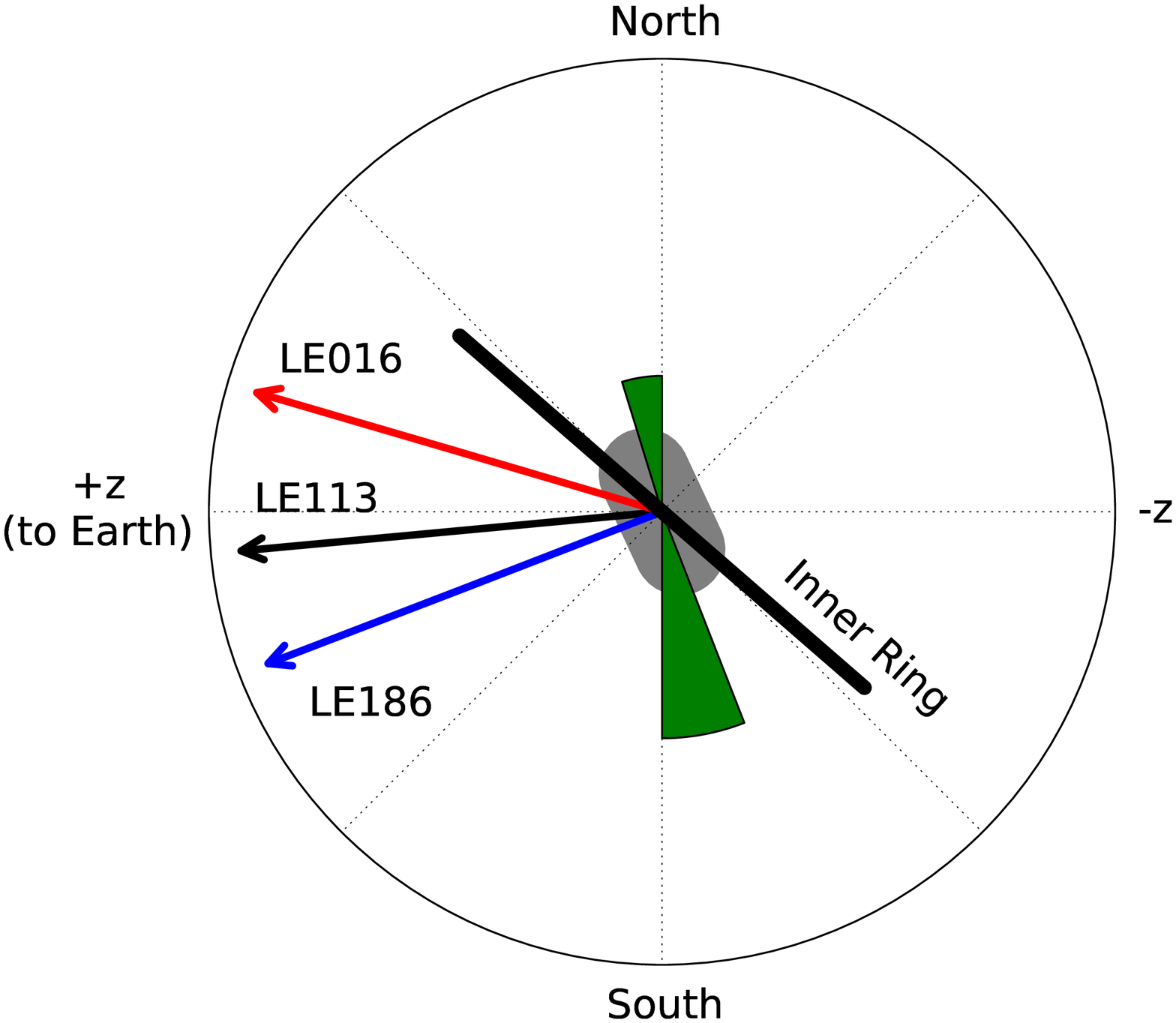} 
 \caption{ LEFT: PAs on the sky of the three dominant LE viewing
angles LE016, LE113 and LE186. The central grey region denotes the
orientation on the sky of the elongated remnant ejecta (PA=16\arcdeg)
from \cite{Kjaer10}.  The green wedges illustrate the proposed
two-sided distribution of $^{56}$Ni, most dominant in the southern
hemisphere. White circles illustrate the locations of the two mystery
spots as identified by \citet{Nisenson99}, with radius proportional to
relative brightness in magnitudes of the two sources.  Only the
relative distance from the center of the SNR to the mystery spots is
to scale in the image. RIGHT: Schematic with viewing angle
perpendicular to Earth’s line of sight. The inclination of the inner
circumstellar ring is shown along with the proposed two-sided
distribution of $^{56}$Ni in green. Note that the green wedges are to
highlight the proposed geometry (not absolute velocity) of the
$^{56}$Ni asymmetry probed by the LE spectra, illustrating that the
southern overabundance is most dominant.  This figure and caption is courtesy of
\cite{Sinnott13}.}
   \label{fig:87aejecta}
\end{center}
\end{figure}

In Figure~\ref{fig:87aejecta}, the light paths of the LE186 and LE016
echoes are illustrated and compared to the structure of the SN~1987A ejecta
\citep{Kjaer10}.
Even though the opening angle of the two LEs is only \about
40$\arcdeg$, the differences in the H$\alpha$ lines are quite striking.
\citet{Sinnott13} argues that these differences are caused by a
two-sided and asymmetric $^{56}$Ni distribution in the outer H envelope. 
The symmetry axis
defined by the 16$\arcdeg$/186$\arcdeg$ viewing angles is in excellent
agreement with the PA of the symmetry axis of the elongated ejecta
that was measured to be \about 15$\arcdeg$ \citep{Wang02,Kjaer10}.
\cite{Kjaer10} also found the present-day ejecta to be blueshifted in
the north and redshifted in the south, inclined out of the plane of
the sky by \about 25$\arcdeg$. The two-sided $^{56}$Ni distribution
proposed by \cite{Sinnott13} as an explanation for the asymmetry seen in
the LE observations is therefore roughly aligned with the $\sim$25-year-old
emerging SN remnant both in PA and inclination out of the sky.

\subsection{Spectroscopic Time Series\label{sec:spectimeseries}}

If the scattering dust filament is infinitely thin, then the 
LE is just the projected light curve of the transient. Therefore it
is, in theory, possible to obtain spectra from different epochs of an
event. In practice, there are complications: The LE profile
gets convolved by the finite thickness of the scattering dust
filament and the finite size of their slit and PSF at the time of observation
\citep{Rest11_leprofile,Rest12b_lerev}.  For typical Galactic LEs 
of ancient SNe and typical dust structures, the spatial extent
is on the order of arcseconds and the temporal resolution ranges from
time scales of a week under the most favorable circumstances, to months
if the scattering dust filament is thick and/or unfavorably
inclined. This means that SNe LEs, which have time-scales of
a couple of months, can only be temporally resolved for very thin dust
filaments and under excellent seeing conditions from the ground or
space. However, transients with much longer time scales can be
resolved much easier. One example is the Great Eruption of $\eta$~Car,
which lasted  two decades and showed temporal variability on
time-scales of months. In this case, LE spectra from different epochs
are only marginally affected by dust width, slit and PSF size.

\begin{figure}[b]
\begin{center}
 \includegraphics[width=2.4in]{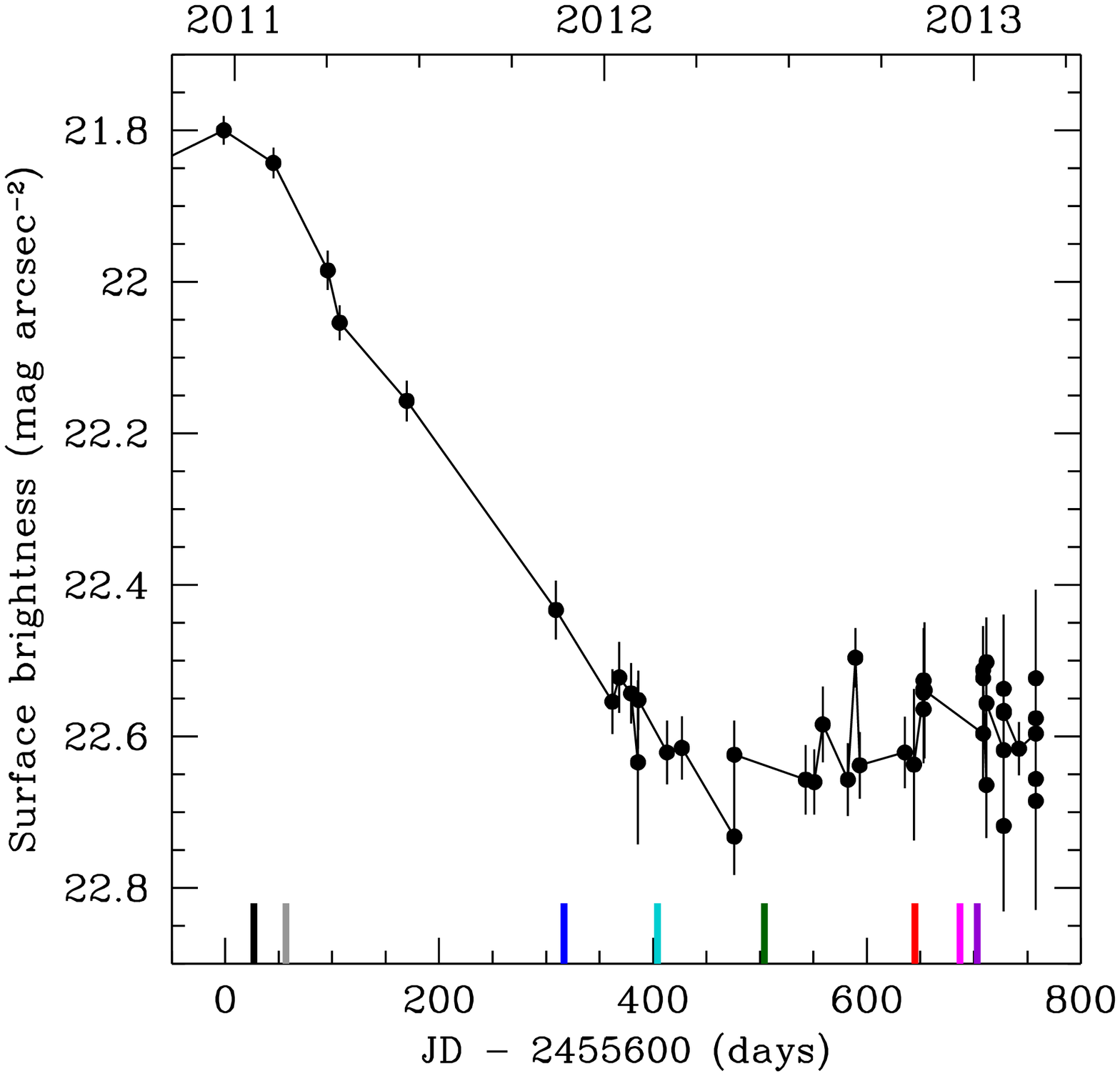} 
 \includegraphics[width=2.4in]{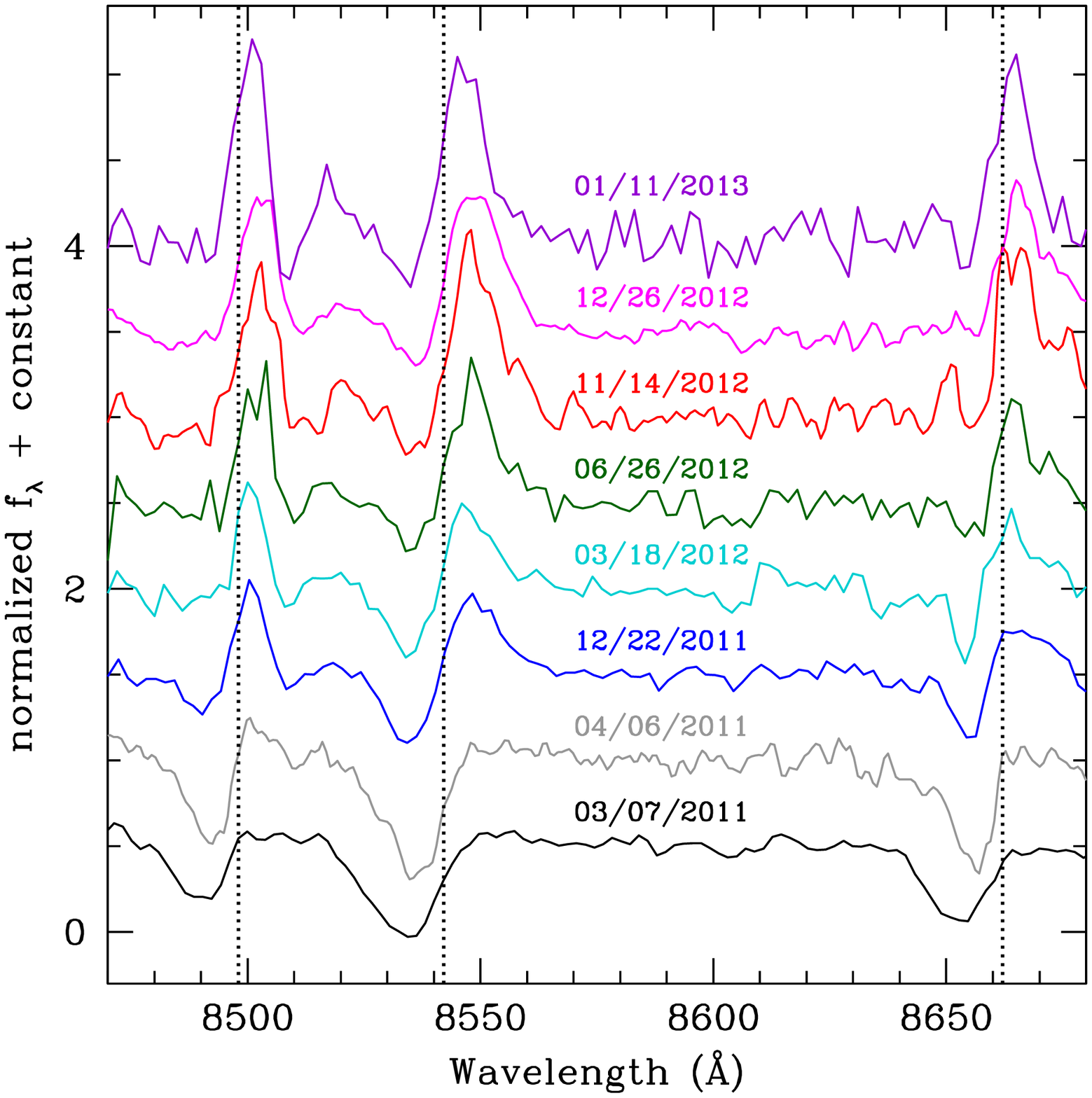} 
 \caption{LEFT: Light curve of part of $\eta$~Car's Great Eruption,
derived from its LEs. Shown is the surface brightness in $i$ band from
the same sky position at different epochs (Blanco 4m (MOSAIC II,
DECam), Swope (Direct CCD), FTS (Spectral), SOAR (SOI)). The LE 
is the projected light curve of the source transient. Since the
LE has an apparent motion, the light curve of the source event
``moves'' through a given position on the sky. The epochs spectra were
taken are indicated with the colored lines.  RIGHT: LE spectra of
$\eta$~Car's Great Eruption showing the Ca~II IR triplet from
different epochs at the same sky location (Magellan Baade (IMACS),
Gemini-S (GMOS)).
} 
  \label{fig:EClcCaII}
\end{center}
\end{figure}

\begin{figure}[b]
\begin{center}
 \includegraphics[width=3.4in]{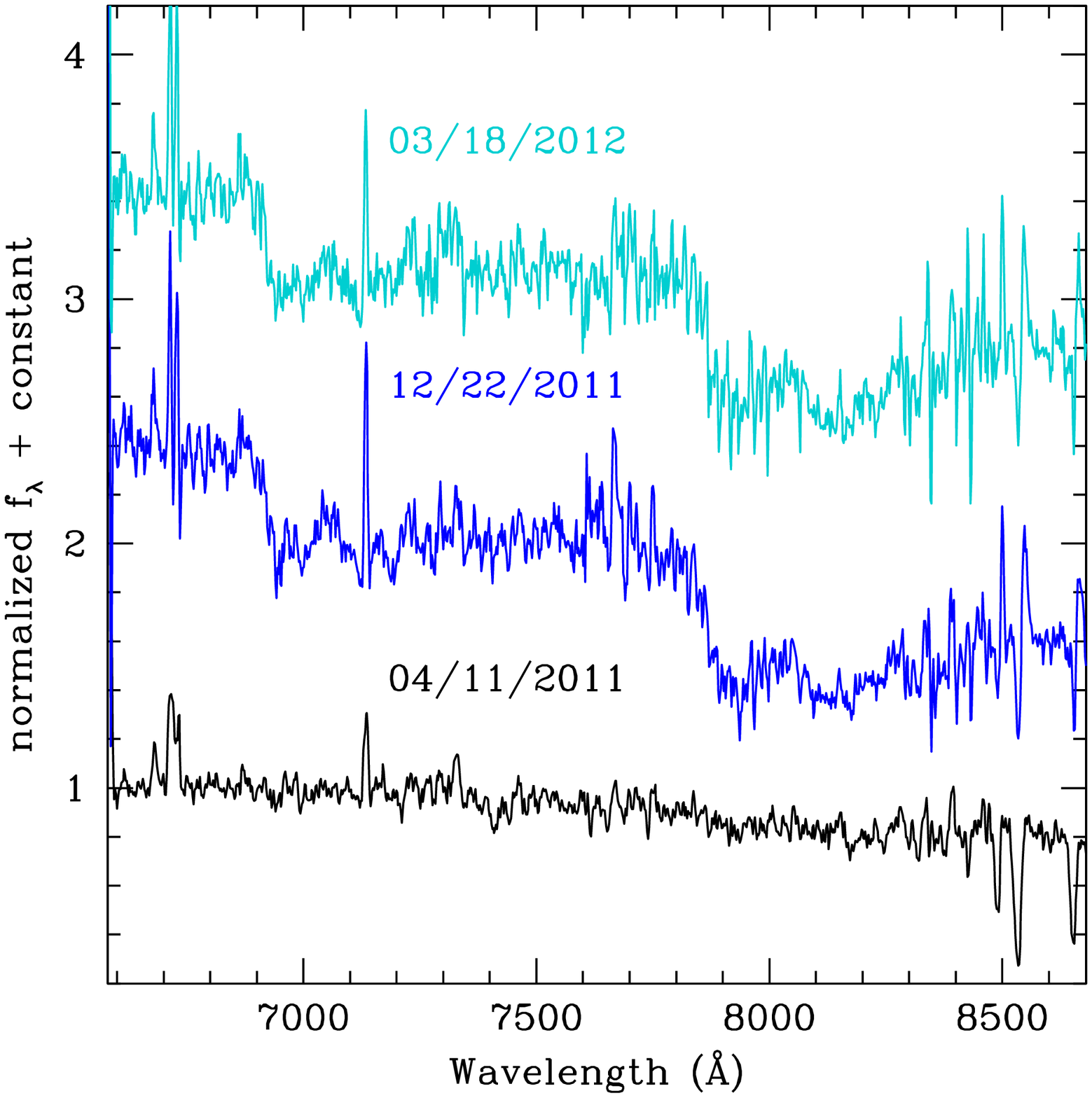} 
 \caption{LE spectra of $\eta$~Car's Great Eruption showing how the CN molecular bands
develop for later epochs.}
   \label{fig:CN}
\end{center}
\end{figure}

The left panel of Figure~\ref{fig:EClcCaII} shows the LE flux
of $\eta$~Car's Great Eruption at a given RA and Dec for various
epochs \citep{Prieto14}. The light curve ``moves'' through a given position, and the
shown flux is the light curve of some part of the Great Eruption
convolved with the scattering dust thickness.
The colored vertical lines indicate when we took a spectrum of the
LE. The right panel of Figure~\ref{fig:EClcCaII} shows the corresponding
spectrum in the same colors for the wavelength range covering the Ca~II
IR triplet \citep{Prieto14}. The spectrum taken close to the peak in the light curve correlates 
best with spectra of G2-to-G5 supergiants, a later range of types than predicted by
standard opaque wind models \citep{Rest12_etac}. The Ca~II IR
triplet is in absorption with an average blueshift of $200$~\kms, and
the lines are asymmetric extending up to blueshifts of $800$~\kms
\citep{Rest12_etac}. Spectra taken at later epochs during the
declining part of the light curve show a transition of the Ca~II IR
triplet from pure absorption through a P-Cygni profile to a nearly
pure emission line spectrum \citep{Prieto14}. In addition, strong CN molecular bands
develop (see Figure~\ref{fig:CN}; \citet{Prieto14}). This is unlike other LBV outbursts
which move back to the earlier stellar types toward the end of the
eruptions. The LEs of $\eta$~Car indicate that the Great
Eruption was not a typical LBV giant eruption.  A paper detailing
these observations is in preparation.

\section{Summary}

In the last decade, LE spectroscopy has emerged as powerful tool to
spectroscopically classify ancient and historic SNe, for which no
contemporary observations with modern instrumentation were possible.
More recently, the technique of LE spectroscopy has been refined and
improved, and it is now possible to utilize it to directly probe the
asymmetries of the transient sources of the LEs. Furthermore, a
spectroscopic time series can be obtained for favorable scattering
dust filament structure or if the source transient is an event with a
long time-scale.

\section{Acknowledgments}

We thank all the observers that have contributed to the monitoring of
$\eta$~Car's LEs, especially E.~Hsiao (and the Carnegie Supernova
Project II) and T.~Matheson.  Based on observations of program
GS-2012B-Q-57, GS-2013A-Q-11, and GS-2013B-Q-19 obtained at the Gemini
Observatory, which is operated by the Association of Universities for
Research in Astronomy, Inc., under a cooperative agreement with the
NSF on behalf of the Gemini partnership: the National Science
Foundation (United States), the National Research Council (Canada),
CONICYT (Chile), the Australian Research Council (Australia),
Minist\'{e}rio da Ci\^{e}ncia, Tecnologia e Inova\c{c}\~{a}o (Brazil)
and Ministerio de Ciencia, Tecnolog\'{i}a e Innovaci\'{o}n Productiva
(Argentina).  This paper includes data gathered with the 6.5 meter
Magellan Telescopes located at Las Campanas Observatory, Chile. Based
on observations at the Cerro Tololo Inter-American Observatory,
National Optical Astronomy Observatory, which are operated by the
Association of Universities for Research in Astronomy, under contract
with the National Science Foundation.  The SOAR Telescope is a joint
project of: Conselho Nacional de Pesquisas Cientificas e Tecnologicas
CNPq-Brazil, The University of North Carolina at Chapel Hill, Michigan
State University, and the National Optical Astronomy Observatory. 
This work was supported by the HST programs GO-12577, AR-12851, and GO-13486.

\bibliographystyle{fapj}
\bibliography{iau296_arest}

\begin{thebibliography}{81}
\expandafter\ifx\csname natexlab\endcsname\relax\def\natexlab#1{#1}\fi

\bibitem[{{Andrews} {et~al.}(2011){Andrews}, {Sugerman}, {Clayton},
  {Gallagher}, {Barlow}, {Clem}, {Ercolano}, {Fabbri}, {Meixner}, {Otsuka},
  {Welch}, \& {Wesson}}]{Andrews11}
{Andrews}, J.~E., {et~al.} 2011, \apj, 731, 47

\bibitem[{{Boffi} {et~al.}(1999){Boffi}, {Sparks}, \& {Macchetto}}]{Boffi99}
{Boffi}, F.~R., {Sparks}, W.~B., \& {Macchetto}, F.~D. 1999, \aaps, 138, 253

\bibitem[{{Bond} {et~al.}(2003){Bond}, {Henden}, {Levay}, {Panagia}, {Sparks},
  {Starrfield}, {Wagner}, {Corradi}, \& {Munari}}]{Bond03}
{Bond}, H.~E., {et~al.} 2003, \nat, 422, 405

\bibitem[{{Boumis} {et~al.}(1998){Boumis}, {Meaburn}, {Bryce}, \&
  {Lopez}}]{Boumis98}
{Boumis}, P., {Meaburn}, J., {Bryce}, M., \& {Lopez}, J.~A. 1998, \mnras, 294,
  61

\bibitem[{{Burrows} {et~al.}(2005){Burrows}, {Walder}, {Ott}, \&
  {Livne}}]{Burrows05}
{Burrows}, A., {Walder}, R., {Ott}, C.~D., \& {Livne}, E. 2005, in Astronomical
  Society of the Pacific Conference Series, Vol. 332, The Fate of the Most
  Massive Stars, ed. {R.~Humphreys \& K.~Stanek}, 350--+

\bibitem[{{Cappellaro} {et~al.}(2001){Cappellaro}, {Patat}, {Mazzali},
  {Benetti}, {Danziger}, {Pastorello}, {Rizzi}, {Salvo}, \&
  {Turatto}}]{Cappellaro01}
{Cappellaro}, E., {et~al.} 2001, \apjl, 549, L215

\bibitem[{{Catchpole} {et~al.}(1987){Catchpole}, {Menzies}, {Monk}, {Wargau},
  {Pollaco}, {Carter}, {Whitelock}, {Marang}, {Laney}, {Balona}, {Feast},
  {Lloyd Evans}, {Sekiguchi}, {Laing}, {Kilkenny}, {Spencer Jones}, {Roberts},
  {Cousins}, {van Vuuren}, \& {Winkler}}]{87aspectra2}
{Catchpole}, R.~M., {et~al.} 1987, \mnras, 229, 15P

\bibitem[{{Catchpole} {et~al.}(1988){Catchpole}, {Whitelock}, {Feast},
  {Menzies}, {Glass}, {Marang}, {Laing}, {Spencer Jones}, {Roberts}, {Balona},
  {Carter}, {Laney}, {Evans}, {Sekiguchi}, {Hutchinson}, {Maddison},
  {Albinson}, {Evans}, {Allen}, {Winkler}, {Fairall}, {Corbally}, {Davies}, \&
  {Parker}}]{87aspectra3}
------. 1988, \mnras, 231, 75P

\bibitem[{{Catchpole} {et~al.}(1989){Catchpole}, {Whitelock}, {Menzies},
  {Feast}, {Marang}, {Sekiguchi}, {van Wyk}, {Roberts}, {Balona}, {Egan},
  {Carter}, {Laney}, {Laing}, {Spencer Jones}, {Glass}, {Winkler}, {Fairall},
  {Lloyd Evans}, {Cropper}, {Shenton}, {Hill}, {Payne}, {Jones}, {Wargau},
  {Mason}, {Jeffery}, {Hellier}, {Parker}, {Chini}, {James}, {Doyle}, {Butler},
  \& {Bromage}}]{87aspectra5}
------. 1989, \mnras, 237, 55P

\bibitem[{{Chevalier}(1986)}]{Chevalier86}
{Chevalier}, R.~A. 1986, \apj, 308, 225

\bibitem[{{Couch} {et~al.}(1990){Couch}, {Allen}, \& {Malin}}]{Couch90}
{Couch}, W.~J., {Allen}, D.~A., \& {Malin}, D.~F. 1990, \mnras, 242, 555

\bibitem[{{Couderc}(1939)}]{Couderc39}
{Couderc}, P. 1939, Annales d'Astrophysique, 2, 271

\bibitem[{{Crotts}(1988)}]{Crotts88}
{Crotts}, A. 1988, \iaucirc, 4561, 4

\bibitem[{{Crotts} \& {Kunkel}(1991)}]{Crotts91}
{Crotts}, A.~P.~S., \& {Kunkel}, W.~E. 1991, \apjl, 366, L73

\bibitem[{{Crotts} {et~al.}(1995){Crotts}, {Kunkel}, \& {Heathcote}}]{Crotts95}
{Crotts}, A.~P.~S., {Kunkel}, W.~E., \& {Heathcote}, S.~R. 1995, \apj, 438, 724

\bibitem[{{Crotts} \& {Yourdon}(2008)}]{Crotts08}
{Crotts}, A.~P.~S., \& {Yourdon}, D. 2008, \apj, 689, 1186

\bibitem[{{DeLaney} {et~al.}(2010){DeLaney}, {Rudnick}, {Stage}, {Smith},
  {Isensee}, {Rho}, {Allen}, {Gomez}, {Kozasa}, {Reach}, {Davis}, \&
  {Houck}}]{Delaney10}
{DeLaney}, T., {et~al.} 2010, \apj, 725, 2038

\bibitem[{{Gallagher} {et~al.}(2011){Gallagher}, {Clayton}, {Andrews},
  {Sugerman}, {Clem}, {Barlow}, {Ercolano}, {Fabbri}, {Wesson}, {Otsuka}, \&
  {Meixner}}]{Gallagher11}
{Gallagher}, J.~S., {et~al.} 2011, in Bulletin of the American Astronomical
  Society, Vol.~43, American Astronomical Society Meeting Abstracts \#217,
  337.22--+

\bibitem[{{Garnavich} {et~al.}(2001){Garnavich}, {Kirshner}, {Challis}, {Jha},
  {Branch}, {Chevalier}, {Filippenko}, {Li}, {Fransson}, {Lundqvist}, {McCray},
  {Panagia}, {Phillips}, {Pun}, {Sonneborn}, {Schmidt}, {Suntzeff}, {Wheeler},
  \& {Supernova INtensive Study (SINS) Collaboration}}]{Garnavich01}
{Garnavich}, P.~M., {et~al.} 2001, in Bulletin of the American Astronomical
  Society, Vol.~33, American Astronomical Society Meeting Abstracts, 1370

\bibitem[{{Gouiffes} {et~al.}(1988){Gouiffes}, {Rosa}, {Melnick}, {Danziger},
  {Remy}, {Santini}, {Sauvageot}, {Jakobsen}, \& {Ruiz}}]{Gouiffes88}
{Gouiffes}, C., {et~al.} 1988, \aap, 198, L9

\bibitem[{{Hamuy} {et~al.}(1988){Hamuy}, {Suntzeff}, {Gonzalez}, \&
  {Martin}}]{87aphoto_hamuy88}
{Hamuy}, M., {Suntzeff}, N.~B., {Gonzalez}, R., \& {Martin}, G. 1988, \aj, 95,
  63

\bibitem[{{Havlen}(1972)}]{Havlen72}
{Havlen}, R.~J. 1972, \aap, 16, 252

\bibitem[{{Kapteyn}(1902)}]{Kapteyn02}
{Kapteyn}, J.~C. 1902, Astronomische Nachrichten, 157, 201

\bibitem[{{Kj{\ae}r} {et~al.}(2010){Kj{\ae}r}, {Leibundgut}, {Fransson},
  {Jerkstrand}, \& {Spyromilio}}]{Kjaer10}
{Kj{\ae}r}, K., {Leibundgut}, B., {Fransson}, C., {Jerkstrand}, A., \&
  {Spyromilio}, J. 2010, \aap, 517, A51

\bibitem[{{Krause} {et~al.}(2008{\natexlab{a}}){Krause}, {Birkmann}, {Usuda},
  {Hattori}, {Goto}, {Rieke}, \& {Misselt}}]{Krause08a}
{Krause}, O., {Birkmann}, S.~M., {Usuda}, T., {Hattori}, T., {Goto}, M.,
  {Rieke}, G.~H., \& {Misselt}, K.~A. 2008{\natexlab{a}}, Science, 320, 1195

\bibitem[{{Krause} {et~al.}(2008{\natexlab{b}}){Krause}, {Tanaka}, {Usuda},
  {Hattori}, {Goto}, {Birkmann}, \& {Nomoto}}]{Krause08b}
{Krause}, O., {Tanaka}, M., {Usuda}, T., {Hattori}, T., {Goto}, M., {Birkmann},
  S., \& {Nomoto}, K. 2008{\natexlab{b}}, \nat, 456, 617

\bibitem[{{Liu} {et~al.}(2003){Liu}, {Bregman}, \& {Seitzer}}]{Liu03}
{Liu}, J.-F., {Bregman}, J.~N., \& {Seitzer}, P. 2003, \apj, 582, 919

\bibitem[{{Menzies} {et~al.}(1987){Menzies}, {Catchpole}, {van Vuuren},
  {Winkler}, {Laney}, {Whitelock}, {Cousins}, {Carter}, {Marang}, {Lloyd
  Evans}, {Roberts}, {Kilkenny}, {Spencer Jones}, {Sekiguchi}, {Fairall}, \&
  {Wolstencroft}}]{87aspectra1}
{Menzies}, J.~W., {et~al.} 1987, \mnras, 227, 39P

\bibitem[{{Miller} {et~al.}(2010){Miller}, {Smith}, {Li}, {Bloom}, {Chornock},
  {Filippenko}, \& {Prochaska}}]{Miller10}
{Miller}, A.~A., {Smith}, N., {Li}, W., {Bloom}, J.~S., {Chornock}, R.,
  {Filippenko}, A.~V., \& {Prochaska}, J.~X. 2010, \aj, 139, 2218

\bibitem[{{Nisenson} \& {Papaliolios}(1999)}]{Nisenson99}
{Nisenson}, P., \& {Papaliolios}, C. 1999, \apjl, 518, L29

\bibitem[{{Ortiz} {et~al.}(2010){Ortiz}, {Sugerman}, {de La Cueva},
  {Santos-Sanz}, {Duffard}, {Gil-Hutton}, {Melita}, \& {Morales}}]{Ortiz10}
{Ortiz}, J.~L., {Sugerman}, B.~E.~K., {de La Cueva}, I., {Santos-Sanz}, P.,
  {Duffard}, R., {Gil-Hutton}, R., {Melita}, M., \& {Morales}, N. 2010, \aap,
  519, A7+

\bibitem[{{Otsuka} {et~al.}(2012){Otsuka}, {Meixner}, {Panagia}, {Fabbri},
  {Barlow}, {Clayton}, {Gallagher}, {Sugerman}, {Wesson}, {Andrews},
  {Ercolano}, \& {Welch}}]{Otsuka12}
{Otsuka}, M., {et~al.} 2012, \apj, 744, 26

\bibitem[{{Patat}(2005)}]{Patat05}
{Patat}, F. 2005, \mnras, 357, 1161

\bibitem[{{Patat} {et~al.}(2006){Patat}, {Benetti}, {Cappellaro}, \&
  {Turatto}}]{Patat06}
{Patat}, F., {Benetti}, S., {Cappellaro}, E., \& {Turatto}, M. 2006, \mnras,
  369, 1949

\bibitem[{{Perrine}(1903)}]{Perrine03}
{Perrine}, C.~D. 1903, \apj, 17, 310

\bibitem[{{Phillips} {et~al.}(1990){Phillips}, {Hamuy}, {Heathcote},
  {Suntzeff}, \& {Kirhakos}}]{Phillips90}
{Phillips}, M.~M., {Hamuy}, M., {Heathcote}, S.~R., {Suntzeff}, N.~B., \&
  {Kirhakos}, S. 1990, \aj, 99, 1133

\bibitem[{{Phillips} {et~al.}(1988){Phillips}, {Heathcote}, {Hamuy}, \&
  {Navarrete}}]{Phillips88}
{Phillips}, M.~M., {Heathcote}, S.~R., {Hamuy}, M., \& {Navarrete}, M. 1988,
  \aj, 95, 1087

\bibitem[{{Prieto} {et~al.}(2014){Prieto}, {Rest}, {Bianco}, {Matheson},
  {Smith}, {Walborn}, {Hsiao}, {Chornock}, {Paredes Alvarez}, {Campillay},
  {Contreras}, {Gonzalez}, {James}, {Knapp}, {Kunder}, {Margheim}, {Morrell},
  {Phillips}, {Smith}, {Welch}, \& {Zenteno}}]{Prieto14}
{Prieto}, J.~L., {et~al.} 2014, ArXiv e-prints, 1403.7202

\bibitem[{{Quinn} {et~al.}(2006){Quinn}, {Garnavich}, {Li}, {Panagia}, {Riess},
  {Schmidt}, \& {Della Valle}}]{Quinn06}
{Quinn}, J.~L., {Garnavich}, P.~M., {Li}, W., {Panagia}, N., {Riess}, A.,
  {Schmidt}, B.~P., \& {Della Valle}, M. 2006, \apj, 652, 512

\bibitem[{{Rest} {et~al.}(2007){Rest}, {Becker}, {Bergmann}, {Blondin},
  {Challis}, {Clocchiatti}, {Cook}, {Damke}, {Garg}, {Huber}, {Lanning},
  {Matheson}, {Minniti}, {Morelli}, {Nikolaev}, {Olsen}, {Oasterle}, {Pignata},
  {Prieto}, {Smith}, {Stubbs}, {Suntzeff}, {Welch}, {Wood-Vasey}, \&
  {Zenteno}}]{Rest07}
{Rest}, A., {et~al.} 2007, in Bulletin of the American Astronomical Society,
  Vol.~38, Bulletin of the American Astronomical Society, 935

\bibitem[{{Rest} {et~al.}(2011{\natexlab{a}}){Rest}, {Foley}, {Sinnott},
  {Welch}, {Badenes}, {Filippenko}, {Bergmann}, {Bhatti}, {Blondin}, {Challis},
  {Damke}, {Finley}, {Huber}, {Kasen}, {Kirshner}, {Matheson}, {Mazzali},
  {Minniti}, {Nakajima}, {Narayan}, {Olsen}, {Sauer}, {Smith}, \&
  {Suntzeff}}]{Rest11_casaspec}
{Rest}, A., {et~al.} 2011{\natexlab{a}}, \apj, 732, 3

\bibitem[{{Rest} {et~al.}(2008{\natexlab{a}}){Rest}, {Matheson}, {Blondin},
  {Bergmann}, {Welch}, {Suntzeff}, {Smith}, {Olsen}, {Prieto}, {Garg},
  {Challis}, {Stubbs}, {Hicken}, {Modjaz}, {Wood-Vasey}, {Zenteno}, {Damke},
  {Newman}, {Huber}, {Cook}, {Nikolaev}, {Becker}, {Miceli}, {Covarrubias},
  {Morelli}, {Pignata}, {Clocchiatti}, {Minniti}, \& {Foley}}]{Rest08a}
------. 2008{\natexlab{a}}, \apj, 680, 1137

\bibitem[{{Rest} {et~al.}(2012{\natexlab{a}}){Rest}, {Prieto}, {Walborn},
  {Smith}, {Bianco}, {Chornock}, {Welch}, {Howell}, {Huber}, {Foley}, {Fong},
  {Sinnott}, {Bond}, {Smith}, {Toledo}, {Minniti}, \& {Mandel}}]{Rest12_etac}
------. 2012{\natexlab{a}}, \nat, 482, 375

\bibitem[{{Rest} {et~al.}(2012{\natexlab{b}}){Rest}, {Sinnott}, \&
  {Welch}}]{Rest12b_lerev}
{Rest}, A., {Sinnott}, B., \& {Welch}, D.~L. 2012{\natexlab{b}}, \pasa, 29, 466

\bibitem[{{Rest} {et~al.}(2011{\natexlab{b}}){Rest}, {Sinnott}, {Welch},
  {Foley}, {Narayan}, {Mandel}, {Huber}, \& {Blondin}}]{Rest11_leprofile}
{Rest}, A., {Sinnott}, B., {Welch}, D.~L., {Foley}, R.~J., {Narayan}, G.,
  {Mandel}, K., {Huber}, M.~E., \& {Blondin}, S. 2011{\natexlab{b}}, \apj, 732,
  2

\bibitem[{{Rest} {et~al.}(2005{\natexlab{a}}){Rest}, {Stubbs}, {Becker},
  {Miknaitis}, {Miceli}, {Covarrubias}, {Hawley}, {Smith}, {Suntzeff}, {Olsen},
  {Prieto}, {Hiriart}, {Welch}, {Cook}, {Nikolaev}, {Huber}, {Prochtor},
  {Clocchiatti}, {Minniti}, {Garg}, {Challis}, {Keller}, \&
  {Schmidt}}]{Rest05a}
{Rest}, A., {et~al.} 2005{\natexlab{a}}, \apj, 634, 1103

\bibitem[{{Rest} {et~al.}(2005{\natexlab{b}}){Rest}, {Suntzeff}, {Olsen},
  {Prieto}, {Smith}, {Welch}, {Becker}, {Bergmann}, {Clocchiatti}, {Cook},
  {Garg}, {Huber}, {Miknaitis}, {Minniti}, {Nikolaev}, \& {Stubbs}}]{Rest05b}
------. 2005{\natexlab{b}}, \nat, 438, 1132

\bibitem[{{Rest} {et~al.}(2008{\natexlab{b}}){Rest}, {Welch}, {Suntzeff},
  {Oaster}, {Lanning}, {Olsen}, {Smith}, {Becker}, {Bergmann}, {Challis},
  {Clocchiatti}, {Cook}, {Damke}, {Garg}, {Huber}, {Matheson}, {Minniti},
  {Prieto}, \& {Wood-Vasey}}]{Rest08b}
------. 2008{\natexlab{b}}, \apjl, 681, L81

\bibitem[{{Ritchey}(1901{\natexlab{a}})}]{Ritchey01b}
{Ritchey}, G.~W. 1901{\natexlab{a}}, \apj, 14, 293

\bibitem[{{Ritchey}(1901{\natexlab{b}})}]{Ritchey01a}
------. 1901{\natexlab{b}}, \apj, 14, 167

\bibitem[{{Ritchey}(1902)}]{Ritchey02}
------. 1902, \apj, 15, 129

\bibitem[{{Schaefer}(1987)}]{Schaefer87a}
{Schaefer}, B.~E. 1987, \apjl, 323, L47

\bibitem[{{Schaefer}(1988)}]{Schaefer88}
------. 1988, \apj, 327, 347

\bibitem[{{Schmidt} {et~al.}(1994){Schmidt}, {Kirshner}, {Leibundgut}, {Wells},
  {Porter}, {Ruiz-Lapuente}, {Challis}, \& {Filippenko}}]{Schmidt94}
{Schmidt}, B.~P., {Kirshner}, R.~P., {Leibundgut}, B., {Wells}, L.~A.,
  {Porter}, A.~C., {Ruiz-Lapuente}, P., {Challis}, P., \& {Filippenko}, A.~V.
  1994, \apjl, 434, L19

\bibitem[{{Sinnott} {et~al.}(2013){Sinnott}, {Welch}, {Rest}, {Sutherland}, \&
  {Bergmann}}]{Sinnott13}
{Sinnott}, B., {Welch}, D.~L., {Rest}, A., {Sutherland}, P.~G., \& {Bergmann},
  M. 2013, \apj, 767, 45

\bibitem[{{Smith} {et~al.}(2003){Smith}, {Davidson}, {Gull}, {Ishibashi}, \&
  {Hillier}}]{Smith03_eta}
{Smith}, N., {Davidson}, K., {Gull}, T.~R., {Ishibashi}, K., \& {Hillier},
  D.~J. 2003, \apj, 586, 432

\bibitem[{{Sparks} {et~al.}(1999){Sparks}, {Macchetto}, {Panagia}, {Boffi},
  {Branch}, {Hazen}, \& {della Valle}}]{Sparks99}
{Sparks}, W.~B., {Macchetto}, F., {Panagia}, N., {Boffi}, F.~R., {Branch}, D.,
  {Hazen}, M.~L., \& {della Valle}, M. 1999, \apj, 523, 585

\bibitem[{{Sugerman}(2003)}]{Sugerman03}
{Sugerman}, B.~E.~K. 2003, \aj, 126, 1939

\bibitem[{{Sugerman}(2005)}]{Sugerman05}
------. 2005, \apjl, 632, L17

\bibitem[{{Sugerman} {et~al.}(2012){Sugerman}, {Andrews}, {Barlow}, {Clayton},
  {Ercolano}, {Ghavamian}, {Kennicutt}, {Krause}, {Meixner}, \&
  {Otsuka}}]{Sugerman12}
{Sugerman}, B.~E.~K., {et~al.} 2012, \apj, 749, 170

\bibitem[{{Sugerman} \& {Crotts}(2002)}]{Sugerman02}
{Sugerman}, B.~E.~K., \& {Crotts}, A.~P.~S. 2002, \apjl, 581, L97

\bibitem[{{Sugerman} {et~al.}(2005{\natexlab{a}}){Sugerman}, {Crotts},
  {Kunkel}, {Heathcote}, \& {Lawrence}}]{Sugerman05a}
{Sugerman}, B.~E.~K., {Crotts}, A.~P.~S., {Kunkel}, W.~E., {Heathcote}, S.~R.,
  \& {Lawrence}, S.~S. 2005{\natexlab{a}}, \apj, 627, 888

\bibitem[{{Sugerman} {et~al.}(2005{\natexlab{b}}){Sugerman}, {Crotts},
  {Kunkel}, {Heathcote}, \& {Lawrence}}]{Sugerman05b}
------. 2005{\natexlab{b}}, \apjs, 159, 60

\bibitem[{{Suntzeff} {et~al.}(1988{\natexlab{a}}){Suntzeff}, {Hamuy}, {Martin},
  {Gomez}, \& {Gonzalez}}]{87aphoto_suntzeff88}
{Suntzeff}, N.~B., {Hamuy}, M., {Martin}, G., {Gomez}, A., \& {Gonzalez}, R.
  1988{\natexlab{a}}, \aj, 96, 1864

\bibitem[{{Suntzeff} {et~al.}(1988{\natexlab{b}}){Suntzeff}, {Heathcote},
  {Weller}, {Caldwell}, \& {Huchra}}]{Suntzeff88}
{Suntzeff}, N.~B., {Heathcote}, S., {Weller}, W.~G., {Caldwell}, N., \&
  {Huchra}, J.~P. 1988{\natexlab{b}}, \nat, 334, 135

\bibitem[{{Swope}(1940)}]{Swope40}
{Swope}, H.~H. 1940, Harvard College Observatory Bulletin, 913, 11

\bibitem[{{van den Bergh}(1965{\natexlab{a}})}]{vandenBergh65b}
{van den Bergh}, S. 1965{\natexlab{a}}, \aj, 70, 667

\bibitem[{{van den Bergh}(1965{\natexlab{b}})}]{vandenBergh65a}
------. 1965{\natexlab{b}}, \pasp, 77, 269

\bibitem[{{van den Bergh}(1966)}]{vandenBergh66}
------. 1966, \pasp, 78, 74

\bibitem[{{van den Bergh}(1977)}]{vandenBergh77}
------. 1977, \pasp, 89, 637

\bibitem[{{Van Dyk}(2013)}]{VanDyk13}
{Van Dyk}, S.~D. 2013, ArXiv e-prints, 1305.6639

\bibitem[{{Van Dyk} {et~al.}(2006){Van Dyk}, {Li}, \& {Filippenko}}]{VanDyk06}
{Van Dyk}, S.~D., {Li}, W., \& {Filippenko}, A.~V. 2006, \pasp, 118, 351

\bibitem[{{Wang} {et~al.}(2002){Wang}, {Wheeler}, {H{\"o}flich}, {Khokhlov},
  {Baade}, {Branch}, {Challis}, {Filippenko}, {Fransson}, {Garnavich},
  {Kirshner}, {Lundqvist}, {McCray}, {Panagia}, {Pun}, {Phillips}, {Sonneborn},
  \& {Suntzeff}}]{Wang02}
{Wang}, L., {et~al.} 2002, \apj, 579, 671

\bibitem[{{Wang} {et~al.}(2008){Wang}, {Li}, {Filippenko}, {Foley}, {Smith}, \&
  {Wang}}]{Wang08}
{Wang}, X., {Li}, W., {Filippenko}, A.~V., {Foley}, R.~J., {Smith}, N., \&
  {Wang}, L. 2008, \apj, 677, 1060

\bibitem[{{Welch} {et~al.}(2007){Welch}, {Clayton}, {Campbell}, {Barlow},
  {Sugerman}, {Meixner}, \& {Bank}}]{Welch07}
{Welch}, D.~L., {Clayton}, G.~C., {Campbell}, A., {Barlow}, M.~J., {Sugerman},
  B.~E.~K., {Meixner}, M., \& {Bank}, S.~H.~R. 2007, \apj, 669, 525

\bibitem[{{Westerlund}(1961)}]{Westerlund61}
{Westerlund}, B. 1961, \pasp, 73, 72

\bibitem[{{Wheeler} {et~al.}(2008){Wheeler}, {Maund}, \& {Couch}}]{Wheeler08}
{Wheeler}, J.~C., {Maund}, J.~R., \& {Couch}, S.~M. 2008, \apj, 677, 1091

\bibitem[{{Whitelock} {et~al.}(1988){Whitelock}, {Catchpole}, {Menziez},
  {Feast}, {Winkler}, {Marang}, {Glass}, {Balona}, {Egan}, {Carter}, {Roberts},
  {Sekiguchi}, {Laney}, {Lloyd Evans}, {Laing}, {Spencer Jones}, {Fernley},
  {James}, {Fairall}, {Monk}, \& {van Wyk}}]{87aspectra4}
{Whitelock}, P.~A., {et~al.} 1988, \mnras, 234, 5P

\bibitem[{{Xu} {et~al.}(1994){Xu}, {Crotts}, \& {Kunkel}}]{Xu94}
{Xu}, J., {Crotts}, A.~P.~S., \& {Kunkel}, W.~E. 1994, \apj, 435, 274

\bibitem[{{Xu} {et~al.}(1995){Xu}, {Crotts}, \& {Kunkel}}]{Xu95}
------. 1995, \apj, 451, 806

\bibitem[{{Zwicky}(1940)}]{Zwicky40}
{Zwicky}, F. 1940, Reviews of Modern Physics, 12, 66

\end{thebibliography}

\end{document}